%% FOLLOWING LINE CANNOT BE BROKEN BEFORE 80 CHAR

# RESIDUAL MESON-MESON INTERACTION FROM LATTICE GAUGE SIMULATION IN A SIMPLE QED$_{2+1}$ MODEL

J. Canosa and H.R. Fiebig [1]

*Physics Department, F.I.U. University Park, Miami, Florida 33199, U.S.A.*

## Abstract

The residual interaction for a meson-meson system is computed utilizing the cumulant, or cluster, expansion of the momentum-space time correlation matrix. The cumulant expansion serves to define asymptotic, or free, meson-meson operators. The definition of an effective interaction is then based on a comparison of the full (interacting) and the free (noninteracting) time correlation matrices. The proposed method, which may straightforwardly be transcribed to other hadron-hadron systems, here is applied to a simple 2+1 dimensional U(1) lattice gauge model tuned such that it is confining. Fermions are treated in the staggered scheme. The effective interaction exhibits a repulsive core and attraction at intermediate relative distances. These findings are consistent with an earlier study of the same model utilizing Lüscher's method where scattering phase shifts are obtained directly.

[1]Supported in part by NSF grant PHY-9409195.

# 1 Introduction

Since the introduction of the $\pi$-meson by Yukawa [1] boson-exchange models have served to describe the phenomenology of the strong force, in particular the nucleon-nucleon interaction [2]. However, an understanding of the nature of the strong interaction between hadrons from first principles, namely quantum chromodynamics (QCD), still remains a central, but nevertheless elusive, problem in nuclear physics to this date. From the nuclear-physics perspective the existence of a confining phase in QCD is a defining feature since it is responsible for allowing only colour-neutral particles to propagate as asymptotic states. The mutual interaction of those particles is a central subject of nuclear physics.

The physics of hadronic interaction probably is very complex. Apart from confinement, some prominent aspects to consider are the internal structure of the hadrons as determined by the quark-gluon-vacuum dynamics of the field theory, further, quark and gluon exchange, the Pauli principle, sea-quark loops, chiral symmetry, features of the SU(3) gauge group, quark flavors and masses, and more. Very likely, different mechanisms contributing to an effective residual interaction will do so in different dynamical situations. For example, sea-quark loops may dominate the long-range attractive part of the interaction, as those are related to virtual $\pi$-meson propagation in the language of boson-exchange models. At small relative distances quark exchange and gluon degrees of freedom may assume the dominant role.

Clearly, a study of hadronic effective interaction based on first principles needs to implement the nonperturbative aspects of the quantum gauge field theory. This is currently best done using lattice gauge field theory and numerical simulation. The number of attempts to understand the nucleon-nucleon interaction from lattice QCD are few. The work of Rabitsch et. al. [3] employs the finite-temperature formalism, no repulsive core is seen. Further, despite huge computational obstacles, recently progress has been made towards the qualitative understanding of $\pi$-N and N-N scattering lengths in QCD with heavy quarks and gluon degrees of freedom [4, 5]. Since the latter simulation already pushes contemporary numerical resources to their limits, the question arises which avenues can be pursued to study the physics of effective hadronic interaction beyond the zero-momentum limit.

The philosophy of the present work is that one may use other, simpler than QCD, field models to capture the basic physics of the effective interaction. The minimal requirements towards such a lattice gauge model is that it is confining, at least for certain choices of the coupling, and that it contains fermions. The hope is that essential dynamical features pertaining to the effective hadronic interaction will be qualitatively similar in QCD. Another important aspect is the development of theoretical concepts and computational techniques which may then be applied to the QCD case in subsequent simulations.

In a previous work [6] meson-meson scattering phase shifts were obtained in a 2+1 dimensional lattice model with a U(1) gauge group and staggered fermions ($QED_{2+1}$). This model evidently oversimplifies structural details of $QCD_{3+1}$. For instance, due to the availability of only one colour baryons are absent, the hadrons in the theory are mesons. On the other hand $QED_{2+1}$ posesses a minimal set of features which should be present in any model aiming at effective hadronic interactions:

- Confinement, at least for some range of the gauge coupling, which allows the propagation of asymptotic colour-singlet particles.

- Fermions, so that a set of hadron-hadron operators can be constructed.

- Space dimension $d \geq 2$, so that various partial waves, or spin channels, can be studied.

Indeed, the results of the scattering simulation [6] were consistent with the gross features of the N-N interaction (short-range repulsion, intermediate-range attraction). On the other hand, Lüscher's method [7], which was utilized to obtain the phase shifts, possesses inherent ambiguities which make it very difficult to interpret the simulation data. Additional, a-priori, assumptions about the effective interaction are almost a necessity. Furthermore, phase shifts are only obtained for a discrete, finite, and typically very small, set of momenta. Interpolation between those is only possible by repeating the simulation on a variety of different-sized lattices.

It is the purpose of this article to develop a viable alternative to Lüscher's method. We will subsequently describe how an effective hadron-hadron interaction can be defined within the framework of a lattice field theory, and how it can be extracted from a numerical simulation. Although the presentation proceeds within the $QED_{2+1}$ model, generalization to other lattice models is straightforward. It is understood (though not attempted here) that the resulting effective interaction may eventually be employed to calculate scattering phase shifts and other two-hadron properties.

Finally, in judging the numerical quality and shortcomings of the present simulation, the reader should be aware that lattice nuclear physics, which we may loosely define as the "chemistry of colour-neutral objects on the lattice" is subject to a change of scale: The residual interaction energies are only $10^{-2}$–$10^{-3}$ times those of a typical hadron mass. Nevertheless, the proposed method, which involves taking ratios of simulation data, is capable of dealing with the transition from the GeV to the MeV scale, although at reduced numerical quality.

## 2 Lattice Model

As in ref. [6] we use an $L^2 \times T = 24^2 \times 32$ lattice with the compact Wilson action for a U(1) gauge group

$$S_W = \beta \sum_{\Box} (1 - \operatorname{Re} U_{\Box}) \qquad (1)$$

where $U_{\Box} = e^{\Theta_{\Box}}$ and $\Theta_{\Box}$ is the oriented plaquette angle. For the fermions the Kogut-Susskind action

$$S_F = \sum_{x,y;f} \bar{\chi}_f(x) G^{-1}_{x,y}[U] \chi_f(y) \qquad (2)$$

is employed [8, 9]. The Grassmann fields $\chi$ and $\bar{\chi}$ have been given an external flavour index $f = u, d$ and

$$G^{-1}_{x,y} = \frac{1}{2} \sum_\mu \eta_\mu(x) \left[ U_\mu(x) \delta_{x+\mu,y} - U^\dagger_\mu(y) \delta_{x,y+\mu} \right] + m_F \delta_{x,y} \tag{3}$$

is the staggered fermion matrix with link variables $U_\mu(x)$ and fermion mass $m_F$. Units are such that the lattice constant is $a = 1$. At $\beta = 1.5$ the lattice model is well in the confined phase, see [6] for details.

## 3  Meson Fields

One-meson fields are constructed from boosted local operators

$$\phi_{\vec{p}}(t) = L^{-2} \sum_{\vec{x}} e^{i \vec{p} \cdot \vec{x}} \bar{\chi}_d(\vec{x}, t) \chi_u(\vec{x}, t) \tag{4}$$

where the sum extends over the $L^2$ sites of the spacial lattice. Thus $\phi_{\vec{p}}$ describes a meson (resembling the $\pi^+$) with momentum

$$\vec{p} = \frac{2\pi}{L}(k_1, k_2) \quad \text{where} \quad k_{1,2} = -(\frac{L}{2} - 1) \ldots \frac{L}{2}, \quad \text{even } L. \tag{5}$$

Two-meson fields with total momentum $\vec{P} = 0$ then are

$$\Phi_{\vec{p}}(t) = \phi_{-\vec{p}}(t) \, \phi_{+\vec{p}}(t). \tag{6}$$

Correlations of these operators contain the information about the dynamics of the meson-meson system and, ultimately, the effective residual interaction.

## 4  Time Correlation Matrices

The 2-point correlator, describing the propagation of *one* meson on the lattice, is

$$C^{(2)}_{\vec{p}\vec{q}}(t, t_0) = \langle \phi^\dagger_{\vec{p}}(t) \, \phi_{\vec{q}}(t_0) \rangle - \langle \phi^\dagger_{\vec{p}}(t) \rangle \langle \phi_{\vec{q}}(t_0) \rangle. \tag{7}$$

It can be worked out in terms of contractions between the Grassmann fields, say

$$\ldots \overset{n}{\chi_f}(x) \ldots \overset{n}{\bar{\chi}_{f'}}(x') \ldots = \ldots \delta_{ff'} \overset{n}{G_{xx'}} \ldots \tag{8}$$

where $n$ indicates the partners of the contraction, and $G$ is the inverse fermion matrix, see (3). We obtain

$$C^{(2)}_{\vec{p}\vec{q}}(t, t_0) = L^{-4} \sum_{\vec{x}} \sum_{\vec{y}} e^{-i\vec{p} \cdot \vec{x} + i\vec{q} \cdot \vec{y}} \langle |G_{\vec{x}t, \vec{y}t_0}|^2 \rangle. \tag{9}$$

Assuming translational invariance in the form

$$\langle |G_{\vec{x}t,\vec{y}t_0}|^2 \rangle = \langle |G_{\vec{x}+\vec{a}t,\vec{y}+\vec{a}t_0}|^2 \rangle \tag{10}$$

one of the site sums in (9) can be worked out, rendering the 2-point correlator diagonal

$$C^{(2)}_{\vec{p}\vec{q}}(t,t_0) = \delta_{\vec{p}\vec{q}}\, L^{-2} \sum_{\vec{x}} e^{-i\vec{p}\cdot\vec{x}} \langle |G_{\vec{x}t,\vec{x}_0 t_0}|^2 \rangle. \tag{11}$$

The point $\vec{x}_0$ is arbitrary, with the correlator $C^{(2)}$ being independent of $\vec{x}_0$. We will later describe how $C^{(2)}$ is computed and explain its importance for extracting the effective interaction.

The 4-point correlator describes the propagation of *two* interacting mesons on the lattice

$$C^{(4)}_{\vec{p}\vec{q}}(t,t_0) = \langle \Phi^\dagger_{\vec{p}}(t)\, \Phi_{\vec{q}}(t_0) \rangle - \langle \Phi^\dagger_{\vec{p}}(t) \rangle \langle \Phi_{\vec{q}}(t_0) \rangle. \tag{12}$$

Here $\vec{p}$ and $\vec{q}$ are the *relative* momenta in the meson-meson system. In terms of fermion propagators this correlator reads

$$\begin{aligned}
C^{(4)}_{\vec{p}\vec{q}}(t,t_0) &= L^{-8} \sum_{\vec{x}_1} \sum_{\vec{x}_2} \sum_{\vec{y}_1} \sum_{\vec{y}_2} e^{i\vec{p}\cdot(\vec{x}_2-\vec{x}_1)+i\vec{q}\cdot(\vec{y}_2-\vec{y}_1)} \\
&\quad \Big\langle \Big( G^*_{\vec{x}_2 t,\vec{y}_2 t_0}\, G_{\vec{x}_2 t,\vec{y}_2 t_0}\, G^*_{\vec{x}_1 t,\vec{y}_1 t_0}\, G_{\vec{x}_1 t,\vec{y}_1 t_0} \\
&\quad + G^*_{\vec{x}_1 t,\vec{y}_2 t_0}\, G_{\vec{x}_1 t,\vec{y}_2 t_0}\, G^*_{\vec{x}_2 t,\vec{y}_1 t_0}\, G_{\vec{x}_2 t,\vec{y}_1 t_0} \\
&\quad - G^*_{\vec{x}_2 t,\vec{y}_1 t_0}\, G_{\vec{x}_2 t,\vec{y}_2 t_0}\, G^*_{\vec{x}_1 t,\vec{y}_2 t_0}\, G_{\vec{x}_1 t,\vec{y}_1 t_0} \\
&\quad - G_{\vec{x}_2 t,\vec{y}_1 t_0}\, G^*_{\vec{x}_2 t,\vec{y}_2 t_0}\, G_{\vec{x}_1 t,\vec{y}_2 t_0}\, G^*_{\vec{x}_1 t,\vec{y}_1 t_0} \Big) \Big\rangle
\end{aligned} \tag{13}$$

which we here list for the sake of completeness. More important for the present purpose is a careful analysis of $C^{(4)}$ with regard to the meson-meson effective interaction which is somehow contained in it. Towards this end a diagramatic classification of the various terms contributing to $C^{(4)}$ proves useful. Such a classification arises naturally from working out the contractions between the quark fields in (12),(6),(4). Let us write

$$C^{(4)} = C^{(4A)} + C^{(4B)} - C^{(4C)} - C^{(4D)} \tag{14}$$

$$= \| \| + \mathbb{X} - \bowtie - \bowtie \tag{15}$$

for the four terms as they occur in (13). Each diagram line corresponds to a fermion propagator. For example, using the notation introduced in (8), we have

$$C^{(4A)} = \langle \overset{43}{\phi^\dagger_{\vec{p}}} \overset{21}{\phi^\dagger_{-\vec{p}}} \overset{12}{\phi_{-\vec{q}}} \overset{34}{\phi_{\vec{q}}} \rangle = \langle \overset{43}{\phi^\dagger_{\vec{p}}} \overset{34}{\phi_{\vec{q}}} \overset{21}{\phi^\dagger_{-\vec{p}}} \overset{12}{\phi_{-\vec{q}}} \rangle \tag{16}$$

$$C^{(4B)} = \langle \overset{21}{\phi^\dagger_{\vec{p}}} \overset{43}{\phi^\dagger_{-\vec{p}}} \overset{12}{\phi_{-\vec{q}}} \overset{34}{\phi_{\vec{q}}} \rangle = \langle \overset{43}{\phi^\dagger_{-\vec{p}}} \overset{34}{\phi_{\vec{q}}} \overset{21}{\phi^\dagger_{\vec{p}}} \overset{12}{\phi_{-\vec{q}}} \rangle \tag{17}$$

where the $n = 1\ldots 4$ identify the partners $\chi$ and $\bar{\chi}$ of a contraction, see (4) and (8). Clearly, in diagrams A and B all pairs of contractions preserve the integrity of the one-meson fields $\phi$ and $\phi^\dagger$ in the sense that no breakup via quark exchange occurs. At the most, diagram B describes the exchange of (composite) mesons as a whole. As we proceed, diagrams A and B will have to be treated on the same footing since the desired effective interaction needs to respect Bose symmetry on the meson level.

Diagrams C and D on the other hand both contain quark exchange between the mesons

$$C^{(4C)} = \langle \overset{23}{\phi^\dagger_{\vec{p}}} \overset{41}{\phi^\dagger_{-\vec{p}}} \overset{12}{\phi_{-\vec{q}}} \overset{34}{\phi_{\vec{q}}} \rangle \tag{18}$$

$$C^{(4D)} = \langle \overset{41}{\phi^\dagger_{\vec{p}}} \overset{23}{\phi^\dagger_{-\vec{p}}} \overset{12}{\phi_{-\vec{q}}} \overset{34}{\phi_{\vec{q}}} \rangle \tag{19}$$

for example $\overset{41}{\phi^\dagger} \ldots \overset{1}{\phi}\overset{4}{\phi}$ etc. Thus these diagrams must be considered solely as sources of the effective interaction.

It should be emphasized that the propagation of free, noninteracting, mesons is not quite described by $C^{(4A)} + C^{(4B)}$. The reason is that gluonic correlations are introduced between the mesons by taking the gauge configuration average $\langle \ldots \rangle$ over the product of all four operators. (This is of course also true for $C^{(4C)}$ and $C^{(4D)}$.) Nevertheless, it is clear that the uncorrelated propagation of two mesons on the lattice is fully contained in $C^{(4A)} + C^{(4B)}$. It is essential to isolate the uncorrelated component because it identifies the channel, in the sense of scattering theory, relative to which an effective interaction can be defined.

## 5 Uncorrelated Mesons

In order to identify the free, noninteracting, part of the meson-meson time correlation matrix let us, for example, examine $C^{(4A)}$. Expressing the contractions in (16) in terms of quark propagators, with the notation introduced in (8), we have

$$C^{(4A)} \sim \langle \overset{4}{G^*} \overset{3}{G} \overset{2}{G^*} \overset{1}{G} \rangle \tag{20}$$

where the $\sim$ indicates the Fourier sums, etc., which carry over from (4). For full mathematical detail see (13).

The gauge configuration average in (20) may be analysed by means of the cumulant (or cluster) expansion theorem, see for example section 12.3 of ref. [10]. Taking advantage of $\langle G \rangle = 0$ we have

$$\langle \overset{4}{G^*} \overset{3}{G} \overset{2}{G^*} \overset{1}{G} \rangle = \langle \overset{4}{G^*} \overset{3}{G} \rangle \langle \overset{2}{G^*} \overset{1}{G} \rangle + \langle \overset{4}{G^*} \overset{1}{G} \rangle \langle \overset{2}{G} \overset{3}{G^*} \rangle + \langle \overset{4}{G^*} \overset{2}{G^*} \rangle \langle \overset{3}{G} \overset{1}{G} \rangle + \langle\!\langle \overset{4}{G^*} \overset{3}{G} \overset{2}{G^*} \overset{1}{G} \rangle\!\rangle . \tag{21}$$

The last term defines the cumulant (which generalizes the standard deviation). The first three terms on the right-hand side of (21) are illustrated in Fig. 1. The dashed lines indicate that the propagators are correlated through gluons. Evidently only the first one of the three separable terms in (21) represents free, uncorrelated, mesons. All

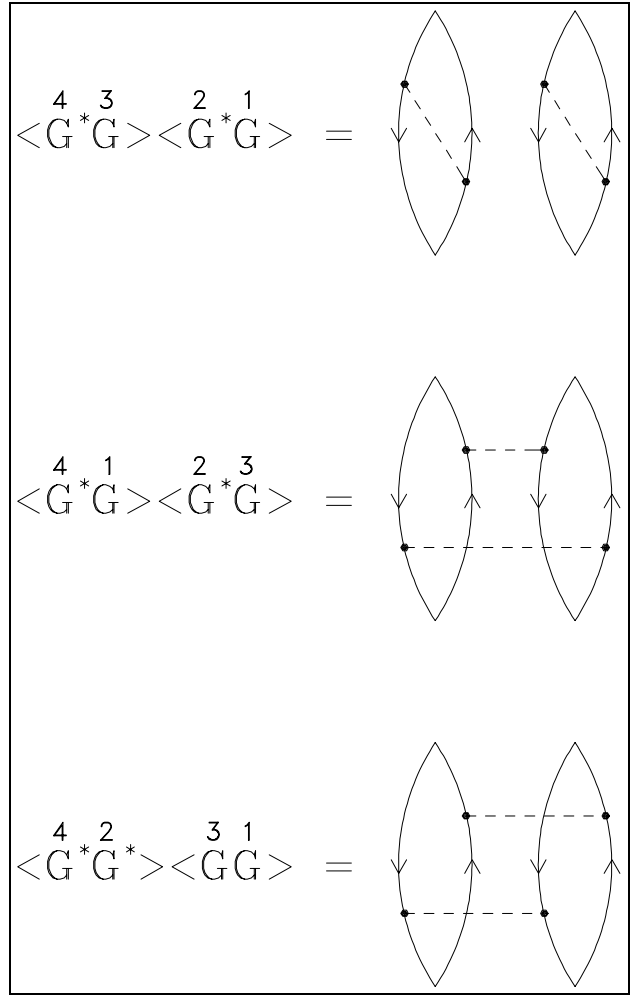

Figure 1: Illustration of the cumulant expansion (21) of the 4-point correlator.

other terms in (21) are sources of residual effective interaction between the mesons. We therefore define

$$\bar{C}^{(4A)} = \langle \phi^\dagger_{\vec{p}} \phi_{\vec{q}} \rangle \langle \phi^\dagger_{-\vec{p}} \phi_{-\vec{q}} \rangle \sim \langle \overset{4}{G}{}^*\overset{3}{G} \rangle \langle \overset{2}{G}{}^*\overset{1}{G} \rangle \, . \tag{22}$$

A similar analysis of $C^{(4B)}$ leads to

$$\bar{C}^{(4B)} = \langle \phi^\dagger_{-\vec{p}} \phi_{\vec{q}} \rangle \langle \phi^\dagger_{\vec{p}} \phi_{-\vec{q}} \rangle \, . \tag{23}$$

The sum of those then is the free meson-meson time correlation matrix

$$\bar{C}^{(4)}_{\vec{p}\vec{q}}(t,t_0) = \bar{C}^{(4A)}_{\vec{p}\vec{q}}(t,t_0) + \bar{C}^{(4B)}_{\vec{p}\vec{q}}(t,t_0) \tag{24}$$

which describes two noninteracting mesons on the lattice. We emphasize that $\bar{C}^{(4)}$ is an additive part of the full 4-point correlator.

$$C^{(4)} = \bar{C}^{(4)} + C^{(4)}_I \tag{25}$$

The remainder $C_I^{(4)}$ contains all sources of the residual interaction, be it from gluonic correlations, or quark exchange (or quark-antiquark loops, if the simluation is unquenched) or any other complicated effects.

It also should be emphasized that the free correlator $\bar{C}^{(4)}$ describes (isolated) mesons the masses of which, as well as other internal structure features, consistently arise from the dynamics determined by the lattice field model and its numerical implementation. From the point of view of numerical simulation it is crucial to capture the small ($10^{-2}$–$10^{-3}$) effects of the residual interaction relative to the computed free correlator.

In the light of the above, the free meson-meson correlator $\bar{C}^{(4)}$ should be obtained numerically from the quark propagator. A glance at (22),(23) and (7) suggests that $\bar{C}^{(4)}$ may be expressed in terms of the 2-point correlator $C^{(2)}$. We note in passing that the separable term in (7) is identically zero because of the $u, d$ flavor assignment to the Grassmann fields of the one-meson operators (4). Thus, using (7) and (22),(23) gives

$$\bar{C}^{(4)}_{\vec{p}\vec{q}} = C^{(2)}_{\vec{p},\vec{q}} C^{(2)}_{-\vec{p},-\vec{q}} + C^{(2)}_{-\vec{p},\vec{q}} C^{(2)}_{\vec{p},-\vec{q}}. \tag{26}$$

Also, recall that $C^{(2)}$ is diagonal in the momentum indices. Writing

$$C^{(2)}_{\vec{p}\vec{q}}(t,t_0) = \delta_{\vec{p}\vec{q}} \, c_{\vec{p}}(t,t_0), \tag{27}$$

where $c_{\vec{p}}(t,t_0)$ is given by comparing (27) and (11), and has the obvious property

$$c_{-\vec{p}}(t,t_0) = c^*_{\vec{p}}(t,t_0), \tag{28}$$

we obtain

$$\bar{C}^{(4)}_{\vec{p}\vec{q}}(t,t_0) = (\delta_{\vec{p},\vec{q}} + \delta_{-\vec{p},\vec{q}}) \, |c_{\vec{p}}(t,t_0)|^2 . \tag{29}$$

In this form $\bar{C}^{(4)}$ is easily computed, for example using (9) to obtain $c_{\vec{p}}(t,t_0) = C^{(2)}_{\vec{p}\vec{p}}(t,t_0)$. This can be done in a manner consistent with the computation of $C^{(4)}_{\vec{p}\vec{q}}(t,t_0)$, for example using (13). Another property, which will become important later, is that the eigenvalues of $\bar{C}^{(4)}$ are proportional to $|c_{\vec{p}}(t,t_0)|^2$ and thus are nonnegative. Finally, it is interesting to note that the property

$$\bar{C}^{(4)}_{\vec{p},\vec{q}} = \bar{C}^{(4)}_{-\vec{p},\vec{q}} = \bar{C}^{(4)}_{\vec{p},-\vec{q}} = \bar{C}^{(4)}_{-\vec{p},-\vec{q}} \tag{30}$$

which is evident from (26) and also holds for $C^{(4)}$, reflects Bose symmetry with respect to the (composite) mesons. Permutation of the mesons, one with momentum $+\vec{p}$ the other with momentum $-\vec{p}$, results in the substitution $\vec{p} \to -\vec{p}$. Clearly the symmetry (30) stems from the fact that $\bar{C}^{(4)}$ was defined through the sum of diagrams $\bar{C}^{(4A)}$ and $\bar{C}^{(4B)}$, see (24).

# 6 Effective Interaction

In principle, the deviation of the full time-correlation matrix $C^{(4)}$ from the free time-correlation matrix $\bar{C}^{(4)}$ contains all the information about the effective meson-meson interaction.

A rigorous definition of the latter emerges from exploring the properties of a time correlation matrix for an elementary interacting meson field subject to canonical quantization. This is discussed in appendix A. In first-order perturbation theory an explicit formula for the (small) interaction can be derived. It involves the zeroth-order correlator, which represents the noninteracting system, and the full correlator, which represents the interacting system. By way of analogy we adopt this formula to guide us towards defining an effective interaction in the framework of a lattice simulation.

Thus, in view of (68), let us define the effective meson-meson time-correlation matrix

$$\mathcal{C}^{(4)}(t,t_0) = \bar{C}^{(4)}(t,t_0)^{-1/2} \, C^{(4)}(t,t_0) \, \bar{C}^{(4)}(t,t_0)^{-1/2} \, . \tag{31}$$

As explained in appendix A multiplying $C^{(4)}$ by the two inverse square roots of $\bar{C}^{(4)}$, in a symmetric way, neutralizes the effects of compositeness of the individual mesons, as well as the trivial aspects of free propagation. The effective correlator $\mathcal{C}^{(4)}$ deviates from the unit matrix (constant for all $t$) only if a relative residual interaction between the mesons is present. The latter then is defined as

$$\mathcal{H}_I = -\left[\frac{\partial \mathcal{C}^{(4)}(t,t_0)}{\partial t}\right]_{t=t_0} \, . \tag{32}$$

This is obviously consistent with

$$\mathcal{C}^{(4)}(t,t_0) = e^{-\mathcal{H}_I(t-t_0)} \, . \tag{33}$$

In appendix A we verify the validity of (the elementary meson field version of) formulas (32) and (33) up to first-order peturbation theory. Equation (33), in conjunction with (31), may simply be adopted as the definition of a residual effective interaction. On the other hand, perturbation theory should be perfectly valid since the effective interaction, as it comes out of a lattice simulation, is about 2-3 orders of magnitude smaller than a typical hadron mass.

# 7 Numerical Procedures

Computation of the 4-point correlation matrix $C^{(4)}$ using the explicit form (13) is numerically not feasible. Not only would it be necessary to compute the entire fermion propagator matrix $G_{\vec{x}t,\vec{y}t_0}$, but also, the four-fold sum over the spacial lattice contains $L^8 \cong 10^{11}$ terms. This computational problem is solved by using a random source technique adapted to the Fourier-type sums which occur in (13). We have used random sources exactly as described in [6], including the choices for all parameters. All correlation matrices were computed in momentum space with a truncation given by

$$\vec{p} = \frac{2\pi}{L}(k_1,k_2) \quad \text{with} \quad |k_{1,2}| \leq 2 \, . \tag{34}$$

The 2-point correlator (which is not needed in Lüscher's method) is obtained from

$$C^{(2)}_{\vec{p},\vec{q}} = \langle \sum_{\vec{x}} e^{-i\vec{p}\cdot\vec{x}} \sum_{<R>} H^*_{\vec{x}t,t_0}(\vec{0};R) H_{\vec{x}t,t_0}(\vec{q};R) \rangle . \tag{35}$$

Here $H_{\vec{x}t,t_0}(\vec{p};R)$ are propagator columns corresponding to Fourier-modified Gaussian random sources exactly as defined in equation (26) of ref. [6], and $\sum_{<R>}$ denotes the average over $N_R = 32$ random sources $R$. Equation (35) is particularly useful for testing the diagonality as claimed in (11). We have verified $C^{(2)}_{\vec{p}\vec{q}}(t,t_0) \propto \delta_{\vec{p}\vec{q}}$ for all momenta allowed by (34), within statistical errors.

Since both the full and the free correlators $C^{(4)}$ and $\bar{C}^{(4)}$ respectively, commute with all elements of the cubic lattice symmetry group $O(d,\mathbf{Z})$ in $d = 2$ (space) dimensions one may compute their reduced matrices, say $C^{(4;\Gamma)}$ and $\bar{C}^{(4;\Gamma)}$, within each irreducible representation $\Gamma$ of $O(d,\mathbf{Z})$. This has been discussed in detail in ref. [6]. We here just mention that there are five irreducible representations $\Gamma$ of $O(d,\mathbf{Z})$, four of which are one-dimensional $\Gamma = A_1, A_2, B_1, B_2$, and one is two-dimensional $\Gamma = E$. The latter is excluded on grounds of the meson-meson system having positive parity. The basis states within each of the former representations spaces $\Gamma$ can be labelled by the magnitude $p = |\vec{p}|$ of the momenta. This is unique for the truncated set (34) of momenta. As is obvious from (29) the reduced free correlator then has the form

$$\bar{C}^{(4;\Gamma)}_{pq}(t,t_0) = \delta_{pq} \left|\bar{c}^{(\Gamma)}_p(t,t_0)\right|^2 \tag{36}$$

for $\Gamma = A_1, A_2, B_1, B_2$, where $\bar{c}^{(\Gamma)}_p(t,t_0)$ are linear combinations of 2-point correlator elements $C^{(2)}_{\vec{p}\vec{p}'}$ with $\vec{p}' = \mathcal{O}_g \vec{p}$, $g \in O(d,\mathbf{Z})$. Those can be easily constructed with the transformations given explicitly in appendix C of ref. [6].

The time behaviour is dominated by an exponential function at large $t$. Since we employ periodic boundary conditions this means

$$\left|\bar{c}^{(\Gamma)}_p(t,t_0)\right|^2 \cong \bar{A}^{(\Gamma)}_p \cosh\left(\bar{W}^{(\Gamma)}_p(t-t_c)\right) \quad \text{for} \quad t \cong t_c \tag{37}$$

with $t_c = T/2 + 1 = 17$, recall $T = 32$. The strength factors $\bar{A}^{(\Gamma)}_p$ and the free two-meson energies $\bar{W}^{(\Gamma)}_p$ are listed in Tab. 1. They were obtained from least-square two-parameter fits to the data using (37). All errors throughout the present work stem from a jackknife procedure [11] with $N_U = 64$ omissions. Some of the momenta $\vec{p}$ belong to more than one sector $\Gamma$, as indicated in Tab. 1. This degeneracy is realized analytically. The strength factors become very small for large relative momenta $p$. This reflects unfavourably on the errors in the last line of Tab. 1. Eventually, data which derive from the largest momentum index, $k^2 = 8$, will be dropped from the subsequent analysis.

Given (36) it is trivial to construct the inverse square roots of the 2-point correlation matrices which occur in the definition (31) of the effective correlator. The reduced matrix elements of the latter thus are

$$\mathcal{C}^{(4;\Gamma)}_{pq}(t,t_0) = \frac{C^{(4;\Gamma)}_{pq}(t,t_0)}{\left|\bar{c}^{(\Gamma)}_p(t,t_0)\right|\left|\bar{c}^{(\Gamma)}_q(t,t_0)\right|} . \tag{38}$$

| $k^2$ | $(k_1,k_2)$ | $\bar{A}_p^{(\Gamma)}\,[5\times 10^{-15}]$ | $\bar{W}_p^{(\Gamma)}$ | Sectors $\Gamma$ | | | |
|---|---|---|---|---|---|---|---|
| 0 | (0,0) | 1.9(2) | 1.04(4) | $A_1$ | | | |
| 1 | (1,0) | 0.38(4) | 1.12(4) | $A_1$ | | $B_1$ | |
| 2 | (1,1) | 0.085(8) | 1.21(4) | $A_1$ | | | $B_2$ |
| 4 | (2,0) | 0.0064(6) | 1.38(6) | $A_1$ | | $B_1$ | |
| 5 | (2,1) | 0.0020(3) | 1.42(5) | $A_1$ | $A_2$ | $B_1$ | $B_2$ |
| 8 | (2,2) | 0.00009(6) | 1.6(5) | $A_1$ | | | $B_2$ |

Table 1: Strength factors and energies for two noninteracting mesons obtained from the free corrrelator $|\bar{c}_p^{(\Gamma)}(t,t_0)|^2$ according to (37). The momenta are referenced using (34).

The effective correlator matrix elements are obtained as a ratio and as such are directly sensitive to the sought-after subtle effects of the residual interaction.

In order to obtain the time derivative of the *effective* correlator matrix $\mathcal{C}^{(4;\Gamma)}(t,t_0)$, see (32), we have first numerically diagonalized the full correlator, say

$$C_{pq}^{(4;\Gamma)}(t,t_0) = \sum_{n=1}^{N^{(\Gamma)}} v_n^{(\Gamma)}(p)\,\lambda_n^{(\Gamma)}(t,t_0)\,v_n^{(\Gamma)}(q) \qquad (39)$$

with $v_n^{(\Gamma)}(p)$ being the components of an orthogonal system of $N^{(\Gamma)}$ real unit vectors $v_n^{(\Gamma)}$ within each of the four allowed sectors $\Gamma$. At large $t$, again meaning $t \cong t_c$, the eigenvalues behave like

$$\lambda_n^{(\Gamma)}(t,t_0) \cong A_n^{(\Gamma)}\cosh\left(W_n^{(\Gamma)}(t-t_c)\right) \quad \text{for} \quad t \cong t_c \qquad (40)$$

whereas the eigenvectors $v_n^{(\Gamma)}$ are independent of $t$. Time-independence of the eigenvectors has been verified numerically for five time slices, $t = 15\ldots 19$, around $t_c = 17$, by computing the statistical errors on the components of the $v_n^{(\Gamma)}$.

The values for $A_n^{(\Gamma)}$ and $W_n^{(\Gamma)}$ in (40) were obtained from two-parameter fits to the data using five time slices $t = 15\ldots 19$. Since this has also been done in the previous work [6], where the same gauge field configuration were used, the results are identical with those in table 2 of ref. [6]. In the sectors $A_1$ and $B_2$ the smallest eigenvalues $\lambda_n^{(\Gamma)}(t,t_0)$ for $n = 6$ and $n = 3$ respectively (which belong to the largest energies) are contaminated by round-off errors beyond control. As mentioned earlier, we face a similar situation in the analysis of the free correlator, as exemplified by the last line of Tab. 1. Thus, our final choice for the sizes of the correlation matrices $C^{(4;\Gamma)}$, $\bar{C}^{(4;\Gamma)}$ and $\mathcal{C}^{(4;\Gamma)}$ was $N^{(\Gamma)} = 5, 1, 3, 2$ for the sectors $\Gamma = A_1, A_2, B_1, B_2$ respectively.

In order to reconcile the cosh time behavior of the correlators with the pure exp behavior assumed in appendix A, and (33), we may choose $t - t_c$ large and negative in (40) and (37), and then drop the negligible ones of the exp terms in the hyperbolic cosines. In this way, from (36)–(40) we obtain

$$C_{pq}^{(4;\Gamma)}(t,t_c) = \sum_{n=1}^{N^{(\Gamma)}} v_n^{(\Gamma)}(p) \frac{A_n^{(\Gamma)}}{\sqrt{\left|\bar{A}_p^{(\Gamma)} \bar{A}_q^{(\Gamma)}\right|}} \exp\left[-\left(W_n^{(\Gamma)} - \frac{\bar{W}_p^{(\Gamma)} + \bar{W}_q^{(\Gamma)}}{2}\right)(t-t_c)\right] v_n^{(\Gamma)}(q). \quad (41)$$

The time derivative in (32), which now has to be taken at $t = t_c$, then leads to

$$\langle p^2 | \mathcal{H}_I^{(\Gamma)} | q^2 \rangle = \sum_{n=1}^{N^{(\Gamma)}} v_n^{(\Gamma)}(p) \frac{A_n^{(\Gamma)}}{\sqrt{\left|\bar{A}_p^{(\Gamma)} \bar{A}_q^{(\Gamma)}\right|}} \left(W_n^{(\Gamma)} - \frac{\bar{W}_p^{(\Gamma)} + \bar{W}_q^{(\Gamma)}}{2}\right) v_n^{(\Gamma)}(q). \quad (42)$$

These are the desired matrix elements of the effective interaction in the basis $|(\Gamma; p^2)\rangle$. From a numerical point of view the computation of those matrix elements is the most critical step of the simulation because it is in (42) where subtle effects of the effective interaction reveal themselves.

To illustrate the numerical situation we have listed in Tab. 2 the components $v_n^{(\Gamma)}(p)$ of the eigenvectors, some strength ratios

$$a_n^{(\Gamma)}(p,q) = \frac{A_n^{(\Gamma)}}{\sqrt{\left|\bar{A}_p^{(\Gamma)} \bar{A}_q^{(\Gamma)}\right|}}, \quad (43)$$

and energy shifts

$$w_n^{(\Gamma)}(p,q) = W_n^{(\Gamma)} - \frac{\bar{W}_p^{(\Gamma)} + \bar{W}_q^{(\Gamma)}}{2}, \quad (44)$$

both for $p = q$, and for the sector $\Gamma = A_1$. As might be expected, given the smallness of the residual interaction, the effective correlator is diagonal to a very high degree. This is reflected in the observation that $v_n^{(\Gamma)}(p) \approx 1$ for only one component of $v_n^{(\Gamma)}$ with all others being quite small. In all cases, however, the largest component of $v_n^{(\Gamma)}$ is multiplied, in (42), with an energy shift $w_n^{(\Gamma)}(p,p)$ of zero, within error limits, see Tab. 2. Also, the strength ratios $a_n^{(\Gamma)}(p,p)$ increase as $p$ becomes larger. This is true for all $n$, however, their overall size is largest for small $n$. It appears that the contributions to the momentum-space matrix elements are mostly driven by the behaviour of the strength ratios, as these vary most dramatically, with their "direction" determined by the signs of the energy shifts. Off-diagonal contributions carry factors $v_n^{(\Gamma)}(p)v_n^{(\Gamma)}(q)$, with $p \neq q$, and are thus somewhat supressed. Those are, however, not negligible. In summary, the numerical situation is best characterized by saying that the matrix elements $\langle p^2 | \mathcal{H}_I^{(\Gamma)} | q^2 \rangle$ emerge in a subtle way from the various factors and terms in (42). The other sectors $\Gamma = A_1, B_1, B_2$ behave similarly, although the magnitude of the factors is significantly smaller.

In Tab. 3 we show the momentum-space matrix elements $\langle p^2 | \mathcal{H}_I^{(\Gamma)} | q^2 \rangle$ for the sector $\Gamma = A_1$. Diagonal elements are numerically zero. The matrix elements are largest in the off-diagonal corners of the matrix.

| $n$ | $k^2$ | $v_n^{(\Gamma)}(p)$ | $a_n^{(\Gamma)}(p,p)$ | $w_n^{(\Gamma)}(p,p)$ |
|---|---|---|---|---|
| 1 | 0 | 0.99899(9) | 0.97(9) | −0.01(4) |
|   | 1 | 0.034(2) | 4.8(4) | −0.09(4) |
|   | 2 | 0.022(2) | 21(2) | −0.18(4) |
|   | 4 | 0.011(2) | 280(30) | −0.35(5) |
|   | 5 | 0.015(2) | 900(100) | −0.39(5) |
| 2 | 0 | −0.037(2) | 0.18(2) | 0.06(4) |
|   | 1 | 0.9963(3) | 0.91(9) | −0.02(4) |
|   | 2 | 0.069(4) | 4.0(4) | −0.11(4) |
|   | 4 | 0.024(3) | 54(5) | −0.29(5) |
|   | 5 | 0.027(4) | 170(30) | −0.32(5) |
| 3 | 0 | −0.020(2) | 0.042(4) | 0.14(4) |
|   | 1 | −0.071(4) | 0.21(2) | 0.06(4) |
|   | 2 | 0.9965(4) | 0.91(8) | −0.03(4) |
|   | 4 | 0.030(7) | 12(1) | −0.20(5) |
|   | 5 | 0.024(9) | 39(6) | −0.24(5) |
| 4 | 0 | −0.009(1) | 0.0035(4) | 0.33(5) |
|   | 1 | −0.022(3) | 0.017(2) | 0.25(5) |
|   | 2 | −0.032(7) | 0.077(8) | 0.16(5) |
|   | 4 | 0.999(1) | 1.0(1) | −0.01(6) |
|   | 5 | −0.02(7) | 3.23(5) | −0.05(6) |
| 5 | 0 | −0.013(1) | 0.0010(2) | 0.38(9) |
|   | 1 | −0.026(3) | 0.0049(9) | 0.30(9) |
|   | 2 | −0.026(9) | 0.022(4) | 0.21(9) |
|   | 4 | 0.02(7) | 0.29(5) | 0.03(9) |
|   | 5 | 0.999(1) | 0.9(2) | 0.00(9) |

Table 2: Selected factors contributing to the matrix elements of the effective interaction according to (42)–(44). The momenta are referenced like in Tab. 1.

|   | 0 | 1 | 2 | 4 | 5 |
|---|---|---|---|---|---|
| 0 | −0.01(4) | 0.004(3) | 0.009(4) | 0.029(8) | 0.08(3) |
| 1 |   | −0.02(4) | 0.009(5) | 0.021(7) | 0.05(2) |
| 2 |   |   | −0.03(4) | 0.004(4) | −0.000(6) |
| 4 |   |   |   | −0.03(6) | −0.05(1) |
| 5 |   |   |   |   | −0.1(1) |

Table 3: Matrix elements $\langle p^2 | \mathcal{H}_I^{(\Gamma)} | q^2 \rangle$ for the sector $\Gamma = A_1$. Row and column indices are labelled by $k^2$, as in Tab. 2.

It should be noted that numerical alternatives to the preceding analysis have been explored. For example, fits to $\bar{c}_p^{(\Gamma)}(t, t_0)$ with a cosh function like (37) lead to one-meson energies which should be close to $\bar{W}_p^{(\Gamma)}/2$. However, given that

$$\cosh[\bar{W}(t - t_c)] = \cosh^2[\frac{\bar{W}}{2}(t - t_c)] + \sinh^2[\frac{\bar{W}}{2}(t - t_c)],$$

this is the case only to some degree of approximation because of contamination from the sinh term. Since the definition of the effective interaction is based on a comparison of two meson-meson correlators, $C^{(4)}$ and $\bar{C}^{(4)}$, it is more consistent to fit to data from $\bar{C}^{(4)}$, as put forward in (36)–(37).

Another alternative is to compute the effective correlator $\mathcal{C}^{(4;\Gamma)}$ directly from the ratios (38), then diagonalize it, and make cosh fits to the eigenvalues with, say

$$\alpha_n^{(\Gamma)} \cosh[\epsilon_n^{(\Gamma)}(t - t_c)].$$

This will yield $\mathcal{H}_I^{(\Gamma)}$ in the diagonal representation, with eigenvalues $\propto \alpha_n^{(\Gamma)}|\epsilon_n^{(\Gamma)}|$. This approach appears very appealing because the ratio data (38) are taken advantage of directly. Unfortunately, the errors on the matrix elements $\mathcal{C}_{pq}^{(4;\Gamma)}(t, t_0)$ tend to be amplified by the diagonalization routine to a larger degree than those of the interacting correlator matrix elements $C_{pq}^{(4;\Gamma)}(t, t_0)$. This effect originates with the ratios (38) for the larger momenta $p$ and $q$ where the numerical values of the numerator and denominator become quite small. The effective interaction obtained in this way appears to be consistent with the results shown below, however, the jackknife analysis of the errors proves difficult because for some of the samples an acceptable (small $\chi^2$) cosh-type fit could not be obtained.

The results presented in the next section are based on employing (42).

# 8 Coordinate-Space Effective Interaction

Given the matrix elements (42) it is straightforward to construct the coordinate-space representation of $\mathcal{H}_I$

$$\langle \vec{r} | \mathcal{H}_I | \vec{r}\,' \rangle = \sum_{\Gamma} \langle \vec{r} | \mathcal{H}_I^{(\Gamma)} | \vec{r}\,' \rangle \tag{45}$$

$$= \sum_{\Gamma} \sum_{p^2, q^2} \langle \vec{r} | (\Gamma; p^2) \rangle \langle p^2 | \mathcal{H}_I^{(\Gamma)} | q^2 \rangle \langle (\Gamma; q^2) | \vec{r}\,' \rangle. \tag{46}$$

The transformation matrices $\langle \vec{r} | (\Gamma; p^2) \rangle$ between the basis $|(\Gamma; p^2)\rangle$ of the sector $\Gamma$ and lattice coordinate points $|\vec{r}\rangle$ are constructed by applying appropriate projection operators to $\langle \vec{r} | \vec{p} \rangle = L^{-1} e^{i\vec{p}\cdot\vec{r}}$. We refer the reader to appendix C of ref. [6].

Of the four representations $\Gamma = A_1, A_2, B_1, B_2$ which contribute to $\langle \vec{r} | \mathcal{H}_I | \vec{r}\,' \rangle$ the $A_1$ sector numerically dominates the $\Gamma$ sum in (45). The $A_1$-sector contribution to the diagonal elements $\langle \vec{r} | \mathcal{H}_I | \vec{r} \rangle$ versus $r = |\vec{r}|$ is shown in Fig. 2. One source of fluctuations

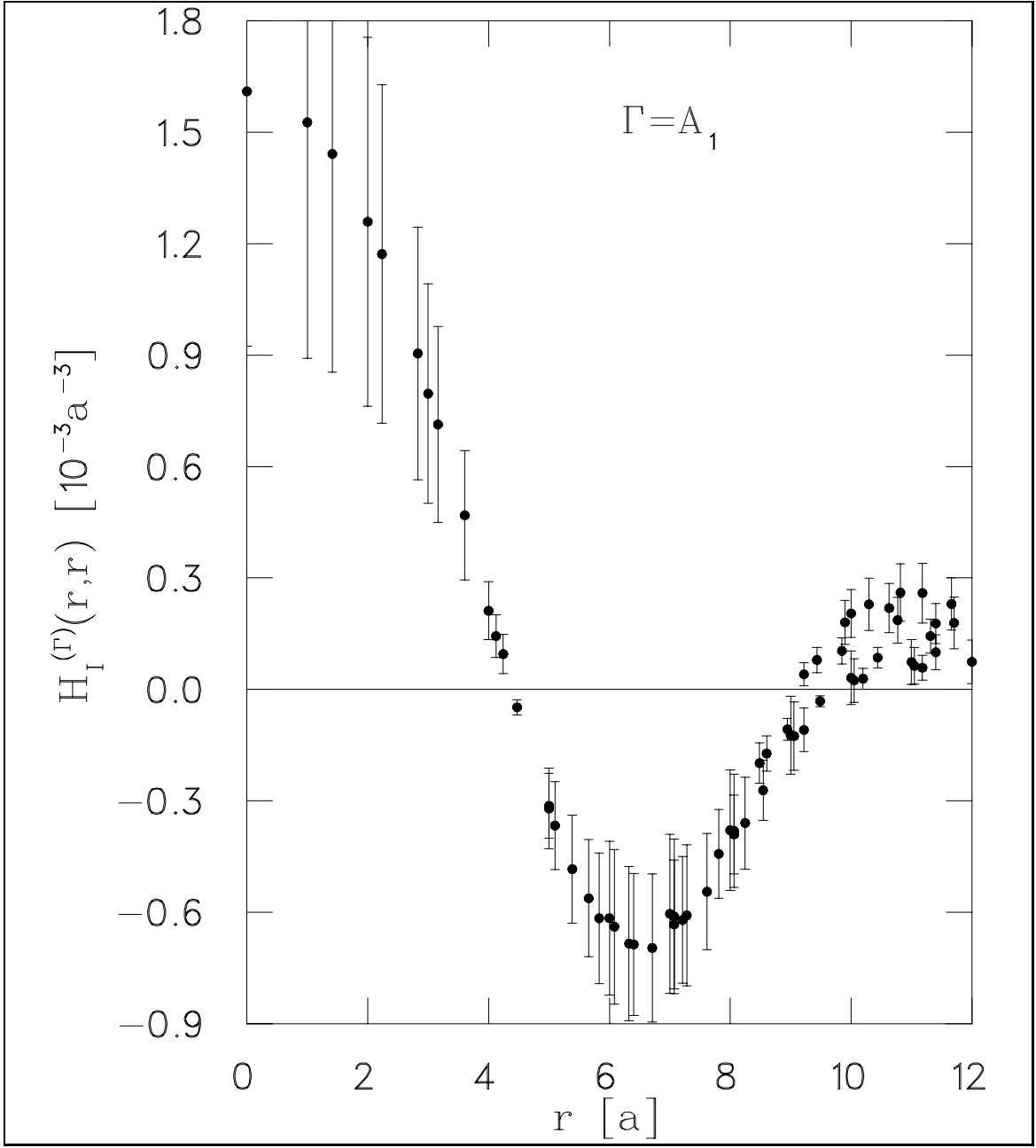

Figure 2: Contribution of the $A_1$ sector to the diagonal matrix elements $\langle \vec{r} | \mathcal{H}_I^{(\Gamma)} | \vec{r} \rangle$ of the effective interaction versus $r = |\vec{r}|$, according to (45).

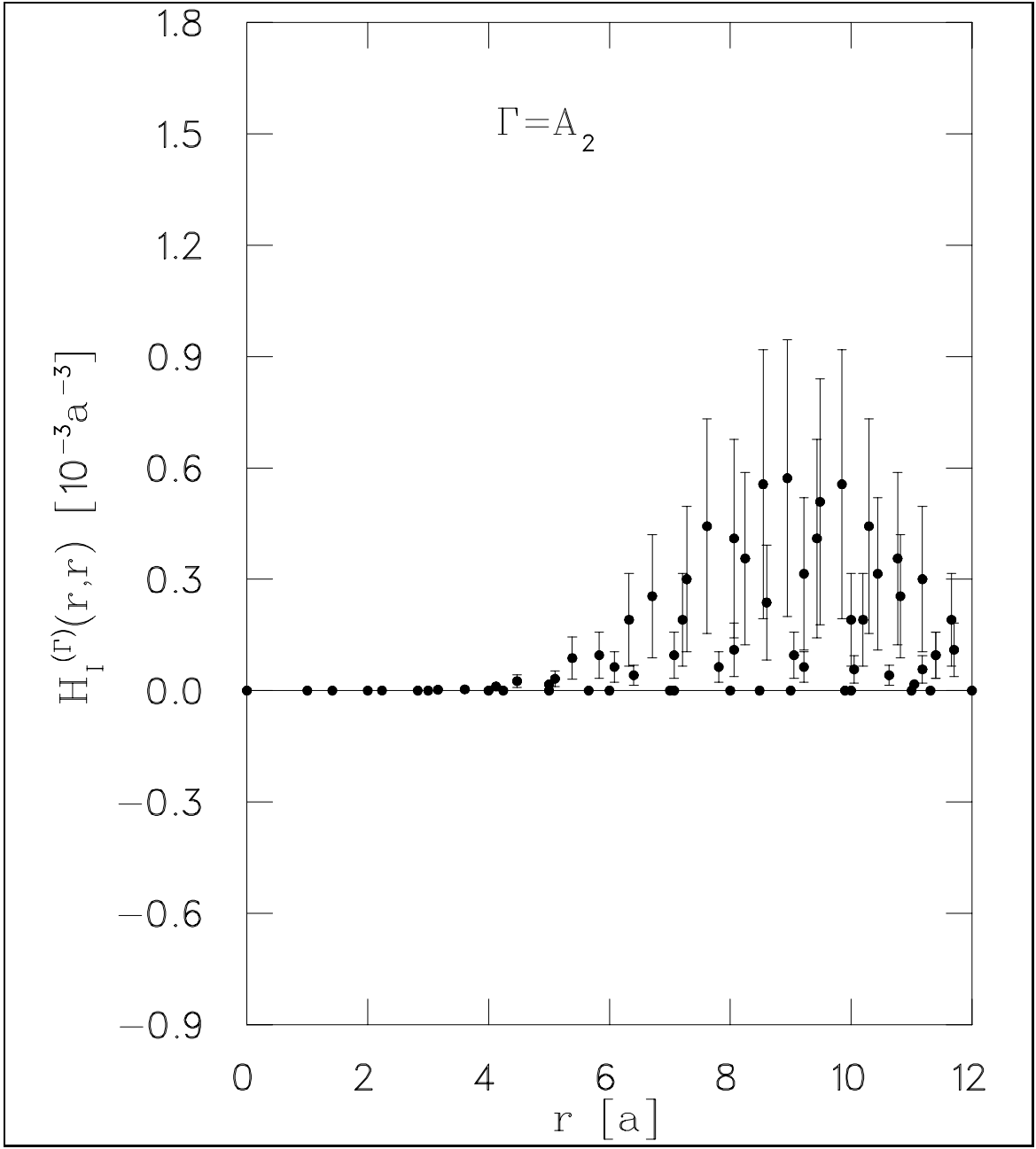

Figure 3: As Fig. 2 for the $A_2$ sector.

of the data points is the lack of continuous rotational symmetry on the lattice. For example, matrix elements $\langle \vec{r} | \mathcal{H}_I^{(\Gamma)} | \vec{r} \rangle$, in $\Gamma = A_1$, are different for $\vec{r} = (5,0)$ and $\vec{r} = (4,3)$, whereas $r = 5$ in both cases. This effect is more pronounced for larger values of $r$, and significantly more in the other sectors $\Gamma$. For illustration Fig. 3 shows the $A_2$ sector contributions to (45).

Since the spacial extent of the lattice is $L = 24$, the maximal relative distance between two mesons, in the presence of periodic boundary conditions, is $r = 12 \times \sqrt{2} \cong 17$ with $\vec{r}$ along a diagonal. However, the data tend to repeat themselves already starting at $r = 12$ which corresponds to $\vec{r}$ aligned along one of the axes of the lattice. Multiple interactions "around the world" between mesons in different copies of the lattice should be expected for $12 \leq r \leq 17$. Interpreting the data in the context of an effective meson-meson interaction is therefore meaningful only in the region $r < 12$.

Statistical errors in Fig. 2 and Fig. 3 stem from a jackknife analysis with $N_U = 64$ omissions. Those matrix elements $\langle p^2 | \mathcal{H}_I^{(\Gamma)} | q^2 \rangle$ which were less in magnitude than their statistical errors have been omitted in the transformation (46).

We have tested the stability of the effective interaction by truncating the matrix in Tab. 3 down to size $4 \times 4$. The coordinate-space effective interaction, which is discussed below, did change by about one $\sigma$, but was still within error limits. It should not be surprising that the qualitative appearance is not changed too much by the truncation since the large-momentum components are only responsible for high-resolution features. The test also shows that the small-momentum approximation to the effective interaction has some degree of justification.

The $A_2$-sector contribution to $\langle \vec{r} | \mathcal{H}_I | \vec{r}' \rangle$ exhibits a very sensitive dependence on the direction of $\vec{r}$, as is evident from Fig. 3. This is also true for the sectors $B_1$ and $B_2$ (not shown), but to a much lesser extent for the $A_1$ sector, see Fig. 2. A sensitive angular dependence is indicative of the presence of nonzero angular momenta. We have used a projection technique, described in appendix B, to interpolate our lattice data for $\langle \vec{r} | \mathcal{H}_I | \vec{r}' \rangle$ with an O(2) rotationally invariant operator. Its partial-wave decomposition is given by the $\ell m$-sum in

$$\langle \vec{r} | \mathcal{H}_I | \vec{r}' \rangle = \sum_{\ell m} Y_{\ell m}(\varphi) \mathcal{H}_I^{(\ell)}(r, r') Y_{\ell m}^*(\varphi') + \ldots \tag{47}$$

with $r = |\vec{r}|$, $r' = |\vec{r}'|$, and $\varphi, \varphi'$ being the polar angles of $\vec{r}$ and $\vec{r}'$ respectively. Here the $Y_{\ell m}(\varphi) \propto e^{im\ell\varphi}$ are (circular) harmonics, as defined in (71) of appendix B. The sum extends over even angular momenta $\ell = 0, 2, 4 \ldots$ only. In appendix B we elaborate on how the partial-wave $\ell$ interpolants $\mathcal{H}_I^{(\ell)}(r, r')$ to the simulation data of the effective interaction are constructed (see (76) for a quick orientation). The ellipses in (47) indicates (small) terms which remain after angular momentum projection, and which in principle should vanish in the continuous limit ($a \to 0$, $L \to \infty$) as the discrete O(2, **Z**) symmetry approaches continuous O(2) symmetry. A good indication of the smallness of those terms, in partial wave $\ell = 0$, is provided by comparing Fig. 4, which shows $\mathcal{H}_I^{(\ell)}(r, r)$ for $\ell = 0$, with the "raw" data of Fig. 2. It appears that "cubic" fluctuations are simply smoothed out. In partial waves $\ell = 2$ and $\ell = 4$ the "raw" data (for example see Fig. 3) are weighted by factors with alternating signs upon projection. This results in very small $\ell = 2$ and $\ell = 4$ wave components for the effective interaction.

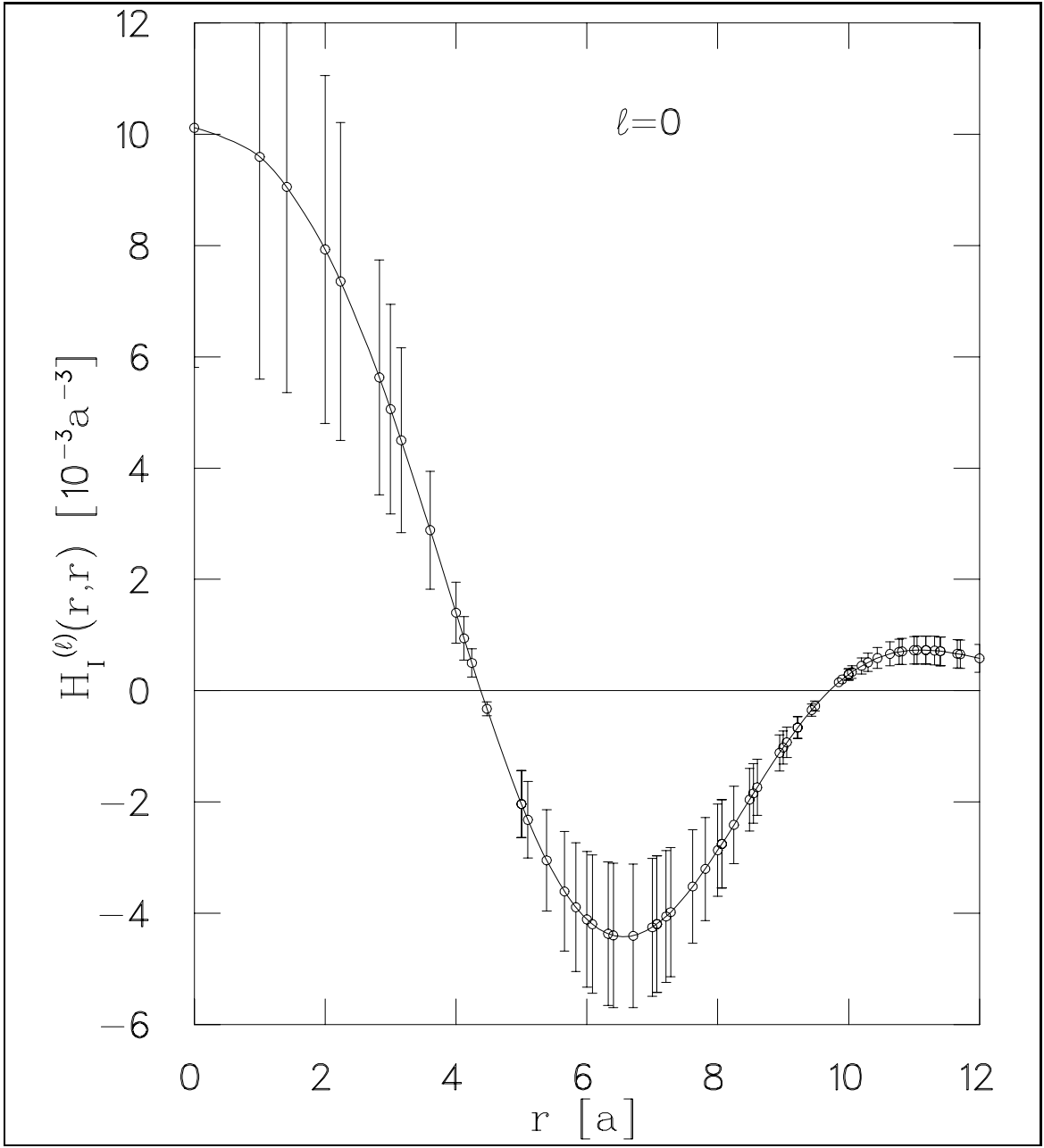

Figure 4: Diagonal elements $\mathcal{H}_I^{(\ell)}(r,r)$ of the angular-momentum-projected effective interaction, for partial wave $\ell = 0$.

As can be seen from (76) and Tab. 4 of appendix B, the partial-wave projected effective interaction $\mathcal{H}_I^{(\ell)}(r,r')$ receives the following contributions:

$\ell = 0$:    $A_1$ only  
$\ell = 2$:    $B_1$ and $B_2$  
$\ell = 4$:    $A_1$ and $A_2$.

The simulation data clearly show that the partial wave $\ell = 0$ dominates over $\ell = 2$ and $\ell = 4$.

The effective meson-meson interaction $\mathcal{H}_I$ obtained from the present lattice simulation is non-local. This is an inevitable consequence of the momentum-space approach used in the present work, see (46). In Figs. 5,6,7 we show 3D plots of $\mathcal{H}_I^{(\ell)}(r,r')$ for $\ell = 0, 2, 4$ respectively.

The salient features of $\mathcal{H}_I^{(\ell=0)}(r,r')$ are repulsion at short distances and a region of attraction at intermediate relative separations. These properties are consistent with earlier results from a scattering phase shift simulation for the same system [6]. For $\ell = 2$ the interaction is much smaller, see Fig. 6, with a repulsive soft core at intermediate $r, r'$ and some attraction at large distances. As $\ell$ increases, the structures are naturally pushed out to larger $r, r'$ as the threshold behavior ($\sim r^\ell$) becomes more dominant. This has apparently happened for $\ell = 4$ to an extent that makes it unfeasible to observe a possible attractive feature at large $r, r'$ because, as mentioned above, $r < 12$ is the upper limit for a valid interpretation of $\mathcal{H}_I$ as a meson-meson interaction.

It is important to be aware of the limitations of the present results. The small-momentum approximation, see (34), naturally limits the resolution of the method with respect to small structures in $\mathcal{H}_I^{(\ell)}(r,r')$. The largest momentum used in the simulation, see Tab. 3, corresponds to a quarter wave length $\lambda/4$ of about 4 lattice units $a$. In particular, the repulsive core, which is a prominent feature in Fig. 5, inevitably appears soft and broad as a result of the Fourier-type approximation. The latter also determines the "visual flavor" of the figures.

It is of some interest to pursue an alternative approach based on a simulation aiming *directly* at coordinate-space matrix elements of $\mathcal{H}_I$. In some sense the latter is a complementary (and incompatible) approximation which has drawbacks of its own, but nevertheless preliminary results [12] seem to indicate a much narrower (hard) repulsive core than Fig. 5 would suggest. A discussion of the latter approach will be forthcoming [13].

In order to examine the effective interaction in terms of, for example, gluon-exchange, quark-exchange, etc., the simulation data will have to be analysed in various other ways. Some of these aspects are currently being targeted by on-going work. However, the origin of the repulsive core (in either of the two approximation schemes) clearly is the anticommuting (Grassmann) nature of the quark fields which imprint certain properties [13] on the effective correlator, at least for the presently studied simple lattice field model and the corresponding class of meson operators.

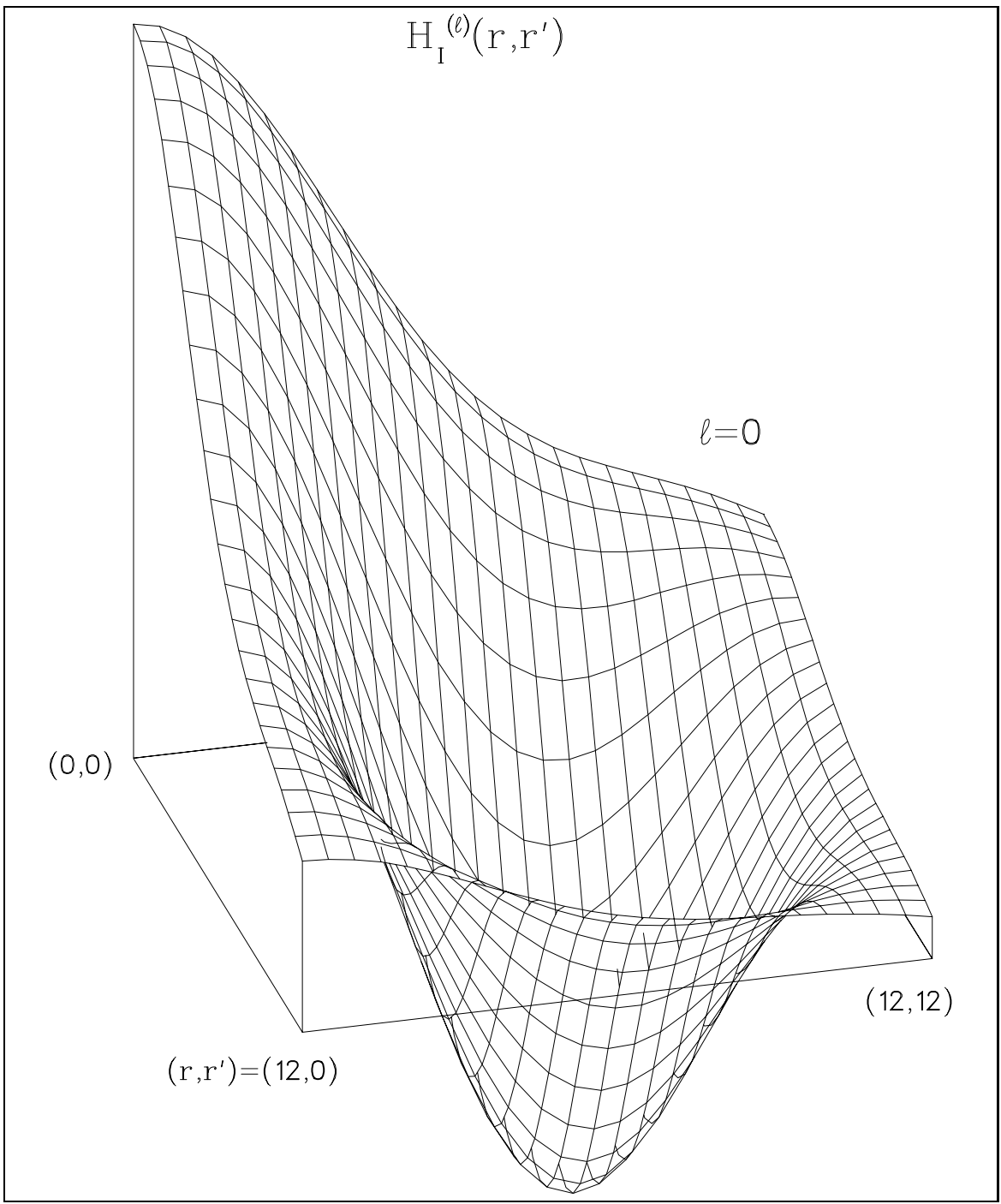

Figure 5: Nonlocal $s$-wave interaction, $\ell = 0$.

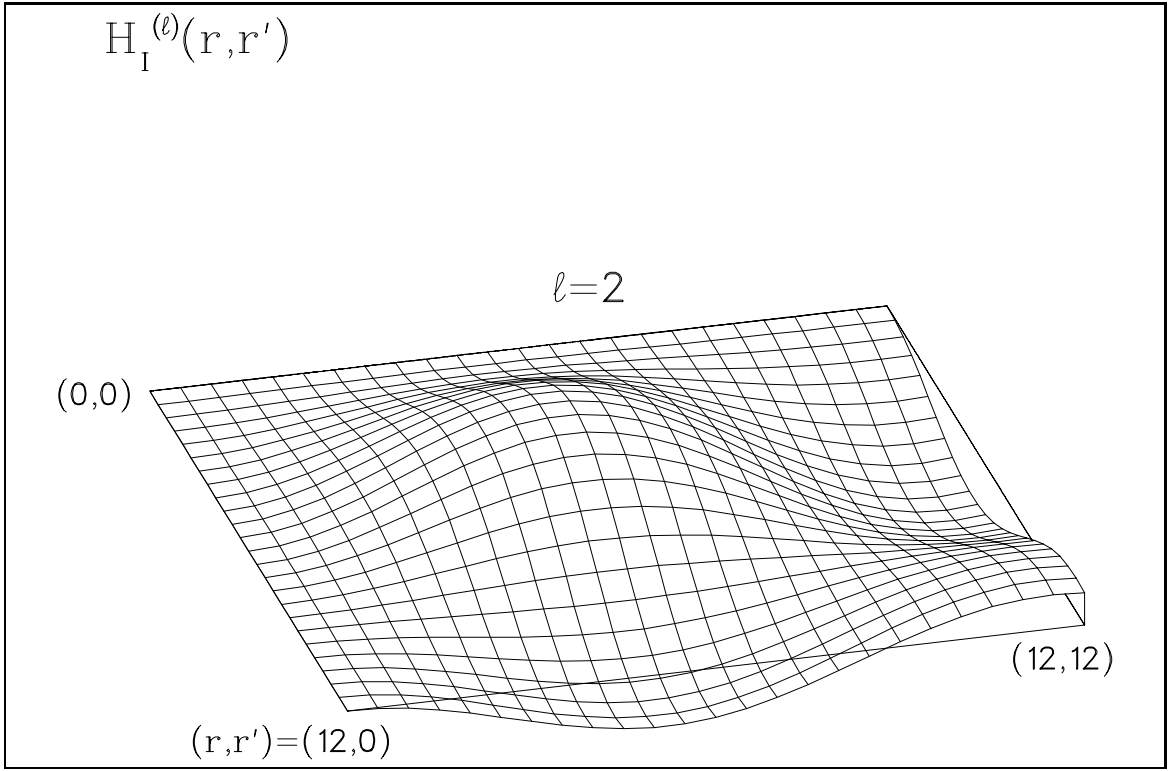

Figure 6: Nonlocal *d*-wave interaction, $\ell = 2$.

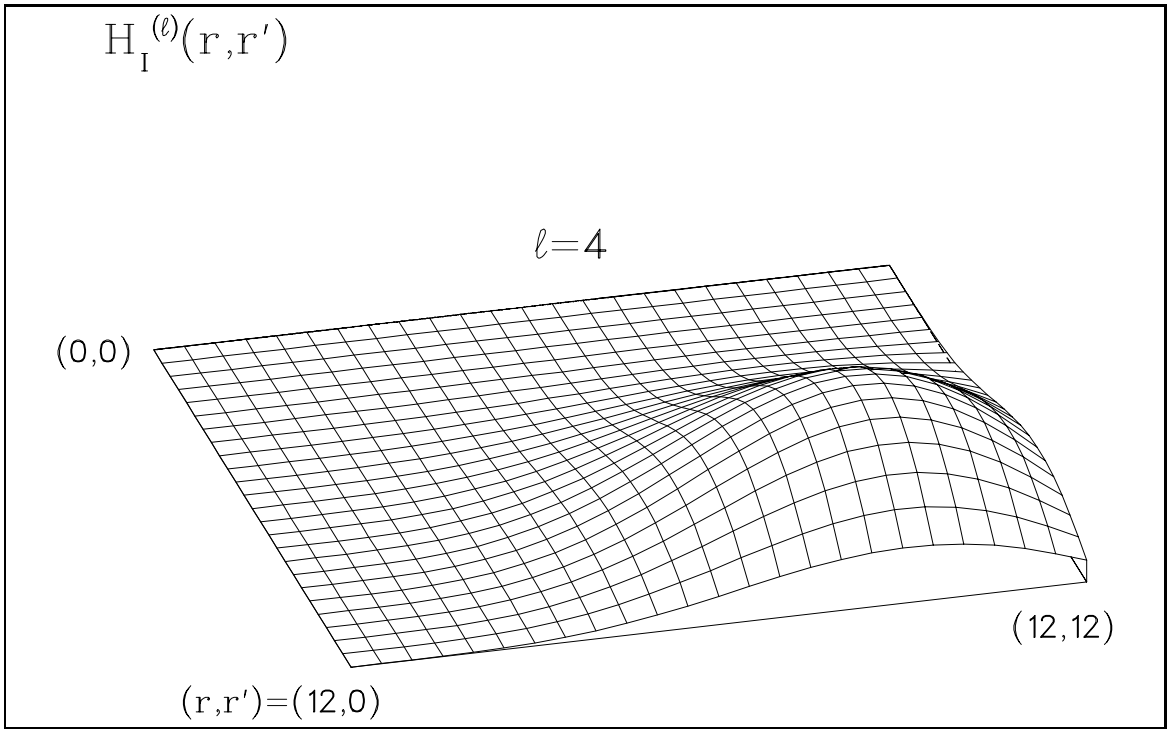

Figure 7: Nonlocal *g*-wave interaction, $\ell = 4$.

# 9  Summary and Conclusion

A method has been developed which allows to extract an effective hadron-hadron interaction from a lattice simulation of a gauge field theory with fermions. Although the practical application has, for now, been limited to a simple field model, a U(1) invariant confining theory in $2 + 1$ dimensions with staggered fermions, the method is eventually intended for studying hadron-hadron interaction in $QCD_{3+1}$. The present proposal also is intended to provide an alternative to Lüscher's approach [7] which suffers from inherent ambiguities related to the interpretation of scattering phase shifts.

In the present application we were able to extract a residual effective meson-meson interaction $\mathcal{H}_I$ which features short-range (Pauli) repulsion and intermediate-range attraction, thus confirming the interpretation of earlier results [6] using Lüscher's approach. It should be emphasized that the effective interaction $\mathcal{H}_I$ is about one to two orders of magnitude smaller than a typical hadron mass. Extracting Mev effects from numbers on a Gev scale represents an extraordinary numerical challenge. Nevertheless, it has been demonstrated that this program is feasible in a lattice simulation, although, naturally, the numerical quality that can be achieved is significantly less than in a standard hadron mass simulation, for example.

In principle the proposed method straightforwardly adapts to applications with $QCD_{3+1}$. Without doubt, problems of numerical magnitude will present themselves. Since it was possible to perform the present $QED_{2+1}$ simulation on a RISC Workstation, there is hope that the use of more powerful computing resources will make simulations of effective hadronic interaction feasible.

It is a pleasure to thank H. Markum for interesting and useful discussions. We also wish to acknowledge the hospitality of CEBAF during the final phase of this work.

# 10 Appendix

## A First-order Perturbative Interaction

We here elaborate on the definition of the effective meson-meson correlation matrix (31) and the effective meson-meson interaction (32) to the extent that those will be derived in first-order perturbation theory for the analogous case of an elementary interacting boson field.

Let $\hat{\mathcal{L}}_0 = \hat{\mathcal{L}}_0(\hat{\phi}, \partial\hat{\phi})$ be the free lagrangian for an elementary boson field $\hat{\phi}(x)$ defined on the sites $x = (\vec{x}, t)$ of the lattice. It is understood that $\hat{\phi}$ is subject to the usual canonical quantization, say through commutators. Be $\hat{\mathcal{L}} = \hat{\mathcal{L}}_0 + \hat{\mathcal{L}}_I$ such that $\hat{\mathcal{L}}_I = \hat{\mathcal{L}}_I(\hat{\phi})$ is a (small) interaction. In the usual way $\hat{\mathcal{L}}$ gives rise to a hamiltonian

$$\hat{H} = \hat{H}_0 + \hat{H}_I \qquad (48)$$

where $\hat{H}_0$ is the free part and $\hat{H}_I$ a perturbative interaction. With view on (4) and (6) define

$$\hat{\phi}_{\vec{p}}(t) = L^{-2} \sum_{\vec{x}} e^{i\vec{p}\cdot\vec{x}} \hat{\phi}(\vec{x}, t) \qquad (49)$$

and

$$\hat{\Phi}_{\vec{p}}(t) = \hat{\phi}_{-\vec{p}}(t)\, \hat{\phi}_{+\vec{p}}(t). \qquad (50)$$

In the correlation matrix

$$\hat{C}^{(4)}_{\vec{p}\vec{q}}(t, t_c) = \langle 0 | \hat{\Phi}^\dagger_{\vec{p}}(t)\, \hat{\Phi}_{\vec{q}}(t_c) | 0 \rangle \qquad (51)$$

the separable term, see (12), has been dropped since it is zero for the flavoured quark fields, see section 4. The nondegenerate vacuum state $|0\rangle$ satisfies $\hat{H}|0\rangle = W_0|0\rangle$. We will assume that its energy is zero, $W_0 = 0$. Thus the time dependence of the correlator (51) may be made explicit

$$\hat{C}^{(4)}_{\vec{p}\vec{q}}(t, t_c) = \langle 0 | \hat{\Phi}^\dagger_{\vec{p}}(t_c)\, e^{-\hat{H}(t-t_c)}\, \hat{\Phi}_{\vec{q}}(t_c) | 0 \rangle. \qquad (52)$$

Switching to the interaction picture, define

$$\hat{H}_I(t) = e^{\hat{H}_0(t-t_c)}\, \hat{H}_I\, e^{-\hat{H}_0(t-t_c)} \qquad (53)$$

and the (euclidean) time evolution operator

$$\hat{U}(t, t_c) = e^{\hat{H}_0(t-t_c)}\, e^{-\hat{H}(t-t_c)}. \qquad (54)$$

The perturbative expansion of the latter

$$\hat{U}(t, t_c) = \sum_{N=0}^{\infty} \frac{(-1)^N}{N!} \int_{t_c}^{t} dt_1 \ldots \int_{t_c}^{t} dt_N\, \mathrm{T}[\hat{H}_I(t_1) \ldots \hat{H}_I(t_N)] \qquad (55)$$

then induces a perturbative expansion of the correlator

$$\hat{C}^{(4)}_{\vec{p}\vec{q}}(t, t_c) = \langle 0|\hat{\Phi}^{\dagger}_{\vec{p}}(t_c)\, e^{-\hat{H}_0(t-t_c)}\, \hat{U}(t, t_c)\, \hat{\Phi}_{\vec{q}}(t_c)|0\rangle \tag{56}$$

$$= \sum_{N=0}^{\infty} \hat{C}^{(4;N)}_{\vec{p}\vec{q}}(t, t_c). \tag{57}$$

The zero-order term $\hat{C}^{(4;N=0)}$ evidently describes noninteracting mesons.

Order $N = 0$

Let $|n\nu\rangle$ be a complete orthogonal set of eigenstates of $\hat{H}_0$

$$\hat{H}_0|n\nu\rangle = W_n^{(0)}|n\nu\rangle \tag{58}$$

where $W_n^{(0)}$ are free two-meson energies on the lattice, and $\nu$ is a degeneracy index. Lattice effects set aside, those energies should be close to $2\sqrt{m^2 + p^2}$ with $m$ being the rest mass of one meson and $p = |\vec{p}|$ its momentum. Also, define the relative meson-meson momentum-space wave functions $\psi^{(0)}_{n\nu}(\vec{p})$ through

$$c^{(0)}_{n\nu}\psi^{(0)}_{n\nu}(\vec{p}) = \langle n\nu|\hat{\Phi}_{\vec{p}}(t_c)|0\rangle^* \tag{59}$$

where $c^{(0)}_{n\nu}$ are normalization factors. The order $N = 0$ correlator then is

$$\hat{C}^{(4;N=0)}_{\vec{p}\vec{q}}(t, t_c) = \sum_{n\nu} |c^{(0)}_{n\nu}|^2 e^{-W_n^{(0)}(t-t_c)} \psi^{(0)}_{n\nu}(\vec{p})\psi^{(0)*}_{n\nu}(\vec{q}). \tag{60}$$

With properly chosen normalization factors $c^{(0)}_{n\nu}$ we expect orthonormality and completeness

$$\sum_{\vec{p}} \psi^{(0)*}_{n\nu}(\vec{p})\psi^{(0)}_{m\mu}(\vec{p}) = \delta_{n\nu,m\mu} \tag{61}$$

$$\sum_{n\nu} \psi^{(0)}_{n\nu}(\vec{p})\psi^{(0)*}_{n\nu}(\vec{q}) = \delta_{\vec{p},\vec{q}}. \tag{62}$$

For a free elementary boson field this is almost a trivial point since the $\psi^{(0)}_{n\nu}$ merely are plane (lattice) waves. A glance at (60) shows that, technically, those could be obtained as (normalized) eigenvectors from diagonalizing the correlation matrix $\hat{C}^{(4;N=0)}$, where $|c^{(0)}_{n\nu}|^2$ are the eigenvalues, at $t = t_c$. For the case considered here all eigenvalues of $\hat{C}^{(4;N=0)}$ will be nonzero.

Considering a lattice model, however, where the role of $\hat{\Phi}_{\vec{p}}(t_c)$ is assumed by a composite operator made from fermion fields, see (4) and (6), it can not be a priori excluded that an operator matrix element, of the type as it occurs in (59), is identically zero for all $\vec{p}$. In this case the free correlator $\bar{C}^{(4)}$ of the lattice model would have an eigenvalue zero (for all $t$). Likewise, if, for some reason, the set of hadron operators used to construct the correlation matrices on the lattice is linearly dependent, one should expect $\bar{C}^{(4)}$ to have an eigenvalue zero.

Order $N = 1$

From (55)–(57) we obtain

$$\hat{C}^{(4;N=1)}_{\vec{p}\vec{q}}(t, t_c) = -\langle 0|\hat{\Phi}^\dagger_{\vec{p}}(t_c) e^{-\hat{H}_0(t-t_c)} \int_{t_c}^t dt_1 \hat{H}_I(t_1) \hat{\Phi}_{\vec{q}}(t_c)|0\rangle . \tag{63}$$

Upon inserting the complete set $|n\nu\rangle$ on both sides of $\hat{H}_I(t_1)$ and using (53) the $t_1$ integral over exponentials can be carried out explicitly. The result is

$$\begin{aligned}
\hat{C}^{(4;N=1)}_{\vec{p}\vec{q}}(t, t_c) &= -\sum_{n\nu}\sum_{m\mu} \psi^{(0)}_{n\nu}(\vec{p})\psi^{(0)*}_{m\mu}(\vec{q}) \\
&\quad \langle n\nu|\hat{H}_I|m\mu\rangle c^{(0)*}_{n\nu} c^{(0)}_{m\mu} \exp\left[-\frac{W^{(0)}_n + W^{(0)}_m}{2}(t-t_c)\right] \\
&\quad \left\{(t-t_c)\delta_{nm} + \frac{\sinh\left[\frac{W^{(0)}_n - W^{(0)}_m}{2}(t-t_c)\right]}{\frac{W^{(0)}_n - W^{(0)}_m}{2}}(1-\delta_{nm})\right\} .
\end{aligned} \tag{64}$$

Without loss of generality the normalization constants $c^{(0)}_{n\nu}$ may be choosen real and positive, with the phase factors being absorbed into $\psi^{(0)}_{n\nu}(\vec{p})$, as is evident from (59). Thus a glance at (60) shows that the two normalization factors and the exponential in (64) may be removed by multiplying the correlation matrix $\hat{C}^{(4;N=1)}$ from both sides with the inverse square root of $\hat{C}^{(4;N=0)}$. Hence the matrix elements of

$$\hat{\mathcal{C}}^{(4;N=1)}(t, t_c) = \hat{C}^{(4;N=0)}(t, t_c)^{-1/2} \hat{C}^{(4;N=1)}(t, t_c) \hat{C}^{(4;N=0)}(t, t_c)^{-1/2} \tag{65}$$

in the basis $\psi^{(0)}_{n\nu}(\vec{p})$ are products of $\langle n\nu|\hat{H}_I|m\mu\rangle$ and the expression inside $\{\ldots\}$ of (64). The $t$ derivative of the latter is equal to one at $t = t_c$. Thus we have

$$\left[\frac{\partial \hat{\mathcal{C}}^{(4;N=1)}_{\vec{p}\vec{q}}(t, t_c)}{\partial t}\right]_{t=t_c} = -\sum_{n\nu}\sum_{m\mu} \psi^{(0)}_{n\nu}(\vec{p})\langle n\nu|\hat{H}_I|m\mu\rangle \psi^{(0)*}_{m\mu}(\vec{q}) . \tag{66}$$

Using (61) and (62), this translates into an explicit equation for $\hat{H}_I$ independent of the basis

$$\hat{H}_I = -\left[\frac{\partial \hat{\mathcal{C}}^{(4;N=1)}(t, t_c)}{\partial t}\right]_{t=t_c} . \tag{67}$$

which is valid to order $N = 1$. Finally, we may replace $\hat{C}^{(4;N=1)}$ in (65) with the full correlation matrix $\hat{C}^{(4)}$. The corresponding expression (67) for $\hat{H}_I$ will still be valid up to order $N = 1$ in perturbation theory. Thus, summarizing, define the effective correlator

$$\hat{\mathcal{C}}^{(4)}(t, t_c) = \hat{C}^{(4;N=0)}(t, t_c)^{-1/2} \hat{C}^{(4)}(t, t_c) \hat{C}^{(4;N=0)}(t, t_c)^{-1/2} \tag{68}$$

understood as a matrix product, then the meson-meson interaction $\hat{H}_I$ satisfies

$$\hat{\mathcal{C}}^{(4)}(t, t_c) = e^{-\hat{H}_I(t-t_c)} \tag{69}$$

in a neighborhood of $t_c$ up to (at least) order $N = 1$ in perturbation theory.

The utility of these results in the framework of a lattice simulation lies in the analogy which can be drawn between $\hat{C}^{(4;N=0)}$ and the free correlator $\bar{C}^{(4)}$, and $\hat{C}^{(4)}$ and the full correlator $C^{(4)}$. The analogue of (69) may then be considered the definition of an effective interaction.

## B Angular Momentum Projection

We here briefly sketch the derivation of the angular momentum-projected matrix elements $\mathcal{H}_I^{(\ell)}(r, r')$ which appear in (47). For additional mathematical detail the reader is referred to ref. [6].

Consider the plane lattice wave $|\vec{p}\rangle$ with $\vec{p} = \frac{2\pi}{L}(k_1, k_2)$, $k_{1,2} = 1 \ldots L$. The coordinate-space components are

$$\langle \vec{r} | \vec{p} \rangle = L^{-1} e^{i\vec{p}\cdot\vec{r}} = \sum_{\ell m} L^{-1} 2\pi i^\ell J_\ell(pr) Y_{\ell m}(\varphi_r) Y_{\ell m}^*(\varphi_p) \,. \tag{70}$$

Here $\vec{r}$ is understood to be a continuous variable, with polar coordinates $r$ and $\varphi_r$, and

$$Y_{\ell m}(\varphi) = \frac{1}{\sqrt{2\pi}} e^{i\ell m \varphi}, \quad \ell = 0, 1 \ldots \infty \,, \tag{71}$$

with $m = 0$ if $\ell = 0$ and $m = \pm 1$ if $\ell > 0$. Now, project out angular momentum $\ell, m$ through

$$\begin{align}
\langle r(\ell m) | \vec{p} \rangle &= \int_0^{2\pi} d\varphi_r \, Y_{\ell m}^*(\varphi_r) \langle \vec{r} | \vec{p} \rangle \tag{72} \\
&= L^{-1} 2\pi i^\ell J_\ell(pr) Y_{\ell m}^*(\varphi_p) \,. \tag{73}
\end{align}$$

This defines states $|r(\ell m)\rangle$ of "good" angular momentum which have support on a discrete set of lattice momenta $\vec{p}$. An operator, say $\mathcal{K}$, with continuous O(2) symmetry would be diagonal in this basis

$$\langle r(\ell m) | \mathcal{K} | s(\lambda \mu) \rangle = \delta_{\ell \lambda} \delta_{m \mu} \mathcal{K}^{(\ell)}(r, s) \,. \tag{74}$$

If $\mathcal{K}$ also has support on the discrete set of momenta $\vec{p}$ we may obtain its coordinate-space partial wave matrix elements from the expansion

$$\mathcal{K}^{(\ell)}(r, s) = \sum_{\Gamma, p^2} \sum_{\Gamma', q^2} \langle r(\ell m) | (\Gamma; p^2) \rangle \langle (\Gamma; p^2) | \mathcal{K} | (\Gamma'; q^2) \rangle \langle (\Gamma'; q^2) | s(\ell m) \rangle \tag{75}$$

where $|(\Gamma; p^2)\rangle$ are the basis states of O(2, **Z**) representations (see beginning of section 8).

Now, assuming $\mathcal{K} = \mathcal{H}_I$, we may take advantage of the O(2, **Z**) invariance of $\mathcal{H}_I$ in (75) and define

$$\mathcal{H}_I^{(\ell)}(r, s) = \sum_\Gamma \sum_{p^2, q^2} \langle r(\ell m) | (\Gamma; p^2) \rangle \langle p^2 | \mathcal{H}_I^{(\Gamma)} | q^2 \rangle \langle (\Gamma; q^2) | s(\ell m) \rangle \,. \tag{76}$$

| (Type) | $\vec{p}$ | $\Gamma$ | $\ell$ | $f_{\ell m}^{(\Gamma)}$ |
|---|---|---|---|---|
| (O) | $p_1 = p_2 = 0$ | $A_1$ | 0 | $1/2$ |
| (P) | $p_1 > 0, p_2 = 0$ | $A_1$ | $4(N-1)$ | 1 |
|  |  | $B_1$ | $4N-2$ | $-1$ |
| (Q) | $p_1 = p_2 > 0$ | $A_1$ | $4(N-1)$ | $(-1)^{N-1}$ |
|  |  | $B_2$ | $4N-2$ | $(-1)^{N-1} i^m$ |
| (G) | $p_1 > p_2 > 0$ | $A_1$ | $4(N-1)$ | $\cos(\ell\gamma)$ |
|  |  | $A_2$ | $4N$ | $-\sqrt{2} i m \sin(\ell\gamma)$ |
|  |  | $A_2$ | $4N$ | $-\sqrt{2} \cos(\ell\gamma)$ |
|  |  | $A_2$ | $4N$ | $\sqrt{2} i m \sin(\ell\gamma)$ |

Table 4: Nonzero factors $f_{\ell m}^{(\Gamma)}$ for the transformation matrix elements (77). The angle $\gamma$ is given by $\tan(\gamma) = p_2/p_1$, and $N \in \mathbf{N}$.

With $\mathcal{H}_I$ failing to be O(2) invariant the diagonality of (74) of course also fails to hold. However, we may now substitute (76) for $\mathcal{K}^{(\ell)}(r,s)$ on the right hand side of (74). This procedure defines the rotationally invariant part of the effective interaction $\mathcal{H}_I$ as obtained from the lattice simulation.

The transformation $\langle r(\ell m)|(\Gamma; p^2)\rangle$ in (76) is easily calculated using (73) and the explicit expressions for $|(\Gamma; p^2)\rangle$ that appear after eqn. (C.17) in appendix C of ref. [6]. All matrix elements have the form

$$\langle r(\ell m)|(\Gamma; p^2)\rangle = \frac{2\pi}{L}\sqrt{\frac{2}{\pi}} J_\ell(pr) f_{\ell m}^{(\Gamma)} . \tag{77}$$

For the Bessel functions $J_\ell(x)$ we use the conventions of [14]. The factors $f_{\ell m}^{(\Gamma)}$ are listed in Tab. 4. Angular momenta are parameterized as $\ell = 4(N-1) = 0, 4, 8 \ldots \infty$ or $\ell = 4N-2 = 2, 6, 10 \ldots \infty$ with $N = 1, 2, 3 \ldots \infty$ in either case. A possible degeneracy of $p^2$, which we do not encounter in the present simulation, is however accounted for in Tab. 4. For all combinations $\Gamma, \ell$ that do not appear in Tab. 4 the corresponding factors $f_{\ell m}^{(\Gamma)}$ are zero.

In (76) a possible $m$ dependence enters only through factors $m^2$ or $im(-i)^m$, which are one in all cases. Thus $m$ disappears completely from (76).